\documentclass[12pt,doublespace]{JHEP3}
\usepackage{amsmath,amssymb,amsthm,amscd,graphicx}
\usepackage{psfrag}
\input epsf.sty
\theoremstyle{definition}

\newcommand{\OO}{{\cal O}}

\newcommand{\CG}{{\cal G}}
\newcommand{\CH}{{\cal H}}

\newcommand{\CN}{{\cal N}}
\newcommand{\CO}{{\cal O}}

\def\IT{{\mathbb T}}
\def\IS{{\mathbb S}}

\newcommand{\tr}{{\rm Tr}}
\newcommand{\re}{{\rm e}}

\newcommand{\rd}{{\rm d}}

\newcommand{\be}{\begin{equation}}
\newcommand{\ee}{\end{equation}}
\newcommand{\ba}{\begin{aligned}}
\newcommand{\ea}{\end{aligned}}
\newcommand{\ben}{\begin{eqnarray}\displaystyle}
\newcommand{\een}{\end{eqnarray}}
\newcommand{\refb}[1]{(\ref{#1})}

\newcommand{\sectiono}[1]{\section{#1}\setcounter{equation}{0}}

\newdimen\tableauside\tableauside=1.0ex
\newdimen\tableaurule\tableaurule=0.4pt
\newdimen\tableaustep
\def\phantomhrule#1{\hbox{\vbox to0pt{\hrule height\tableaurule width#1\vss}}}
\def\phantomvrule#1{\vbox{\hbox to0pt{\vrule width\tableaurule height#1\hss}}}
\def\sqr{\vbox{%
  \phantomhrule\tableaustep
  \hbox{\phantomvrule\tableaustep\kern\tableaustep\phantomvrule\tableaustep}%
  \hbox{\vbox{\phantomhrule\tableauside}\kern-\tableaurule}}}
\def\squares#1{\hbox{\count0=#1\noindent\loop\sqr
  \advance\count0 by-1 \ifnum\count0>0\repeat}}
\def\tableau#1{\vcenter{\offinterlineskip
  \tableaustep=\tableauside\advance\tableaustep by-\tableaurule
  \kern\normallineskip\hbox
    {\kern\normallineskip\vbox
      {\gettableau#1 0 }%
     \kern\normallineskip\kern\tableaurule}%
  \kern\normallineskip\kern\tableaurule}}
\def\gettableau#1{\ifnum#1=0\let\next=\null\else
\squares{#1}\let\next=\gettableau\fi\next}

\title{\Huge{A one-loop test of quantum supergravity}}

\author{
Sayantani Bhattacharyya$^a$, Alba Grassi$^b$, Marcos Mari\~no$^b$ and Ashoke Sen$^a$ \\
$^a$Harish Chandra Research Institute,\\
\phantom{$^a$}%
Chhatnag Road, Jhusi, Allahabad 211019, India\\
$^b$D\'epartement de Physique Th\'eorique et Section de Math\'ematiques,\\
\phantom{$^b$}%
Universit\'e de Gen\`eve, Gen\`eve, CH-1211 Switzerland\\
\\
\email{sayanta@mri.ernet.in}, \quad \email{alba.grassi@unige.ch}, \quad
\email{marcos.marino@unige.ch}, \quad
\email{sen@mri.ernet.in}

}

\abstract{The partition function on the three-sphere of ABJM theory and its generalizations has, at large $N$, a universal, subleading logarithmic term. 
Inspired by the success of one-loop quantum gravity for computing
the logarithmic corrections to black hole entropy, 
we try to reproduce this universal term by a one-loop calculation 
in Euclidean eleven-dimensional supergravity on AdS$_4\times X_7$. 
We find perfect agreement between the results of ABJM theory and 
the eleven dimensional supergravity.
}    

\begin{document}

\sectiono{Introduction}

Recently, various exact results have been obtained on the partition function on the 
three-sphere of supersymmetric Chern--Simons--matter (CSM) theories generalizing ABJM theory \cite{abjm}. Typical examples are quiver $\CN=3$ CSM theories \cite{jt}. 
These theories are parametrized by the rank $N$ of the $U(N)$ gauge group, 
the CS levels $k_a$, and the number of flavors in each node, $N_{f_a}$. 
Here, $a=1,\cdots,p$ is an index labeling the node. For these theories, 
the partition function at all orders in $1/N$ was computed in \cite{fhm,mp} by generalizing earlier work in \cite{0912.3074,1007.3837,1011.5487,dmp-np}.
The result is given by an Airy function 
\be
\label{zairy}
Z_{\rm CFT}\left(N, \{k_a\}, \{N_{f_a}\}\right) \propto {\rm Ai}\left[ 
C\left(\{k_a\}, \{N_{f_a}\}\right)^{-1/3} \left( N -B\left(\{k_a\}, \{N_{f_a}\} \right) \right)\right].
\ee
In this expression, $C(\{k_a\}, \{N_{f_a}\})$ is a known function of the parameters 
$k_a$ and $N_{f_a}$. The function 
$B(\{k_a\}, \{N_{f_a}\})$ can be computed by a precise algorithm in a case-by-case basis \cite{mp}. 
The proportionality coefficient in (\ref{zairy}) is independent of $N$, but it is a non-trivial function of $k_a$ and $N_{f_a}$ (see \cite{mp,hanada} for 
explicit results on this coefficient in ABJM theory). 

If we recall the asymptotic expansion of the Airy function, 
\be
{\rm Ai}(x) \sim{\re^{-{2\over 3} x^{3/2}} \over 2 {\sqrt{\pi}} x^{1/4}}, \qquad x\gg 1, 
\ee
we find the large $N$ expansion of the free energy  $F\equiv \ln Z$
\ben
\label{largenfree}
F_{\rm CFT}\left(N, \{k_a\}, \{N_{f_a}\}\right) &=& -{2\over 3} C\left(\{k_a\}, \{N_{f_a}\}
\right)^{-1/2} N^{3/2} + C\left(\{k_a\}, \{N_{f_a}\}
\right)^{-1/2} B\left(\{k_a\}, \{N_{f_a}\} \right) N^{1/2}\nonumber \\
&& -{1\over 4} \log \, N + {\rm constant}+ 
\CO\left({1\over {\sqrt{N}}}\right).
\een

These CSM theories are conjecturally dual to M-theory compactifications on manifolds of the form AdS$_4\times X_7$, where $X_7$ is a tri-Sasaki Einstein  
manifold whose geometry 
is specified by the data of the quiver $k_a$, $N_{f_a}$. In particular, the function 
$C\left(\{k_a\}, \{N_{f_a}\}\right)$ is given by \cite{mp,1011.5487,herzog}
\be
C\left(\{k_a\}, \{N_{f_a}\}\right)={6 {\rm vol}(X_7) \over \pi^6}\, ,
\ee
where in the definition of ${\rm vol}(X_7)$  
we have taken out a factor of $(2L)^7$,
$L$ being the `radius' of AdS$_4$. With this convention ${\rm vol}(X_7)$
is  a purely numerical factor.
The leading term of the free energy is then 
\be
\label{leadinglargen}
-{\sqrt{ 2 \pi^6 \over 27 {\rm vol}(X_7)}}N^{3/2}.
\ee
If we use the dictionary relating $N$ to the AdS 
radius $L$ 
\be
(2\pi \ell_p)^6 N=6 (2L)^6 {\rm vol}(X_7)  
\ee
it can be shown that (\ref{leadinglargen}) is (minus) the regularized, gravitational action on-shell, as expected from the AdS/CFT correspondence (see for example \cite{mmlectures} 
for a review). This 
provides a non-trivial, quantitative check of the correspondence at leading order in $N$.  

The first subleading correction of order $N^{1/2}\sim L^3$ originates 
from a local 8-derivative correction to the effective action that modifies
the relation between $N$ and $L$ by shifting $N$ by 
$-B\left(\{k_a\}, \{N_{f_a}\} \right)$\cite{0902.1743,0906.2390}. 
A similar shift in the context of five dimensional black hole entropy has been discussed
in \cite{0807.0237}.

What about the subleading, logarithmic correction appearing in (\ref{largenfree})? When expressed in terms of the AdS radius, it leads to a term of the form 
\be
\label{slog}
-{3\over 2} \log \, L. 
\ee
Notice that this term is {\it universal}: it is the same for all compactifications, irrespectively of the $X_7$ manifold. In fact, there is some evidence that, even for theories with $\CN=2$ supersymmetry, 
the partition function is also an Airy function \cite{mp-two}, and therefore one finds the same type of logarithmic correction. 

A natural question is: can one test the result of this ``microscopic," gauge theory computation of the partition function, in terms of a ``macroscopic" computation in AdS gravity? 

Recently, a similar question has been answered in the affirmative in a related context. In string theory there are by now many exact formulae for the microscopic 
entropy of extremal black holes, as a function of the charges. In the limit 
of large charges, this entropy agrees with the Bekenstein--Hawking entropy (or, more generally, with Wald's entropy). However, there are subleading corrections in the asymptotic expansion of the exact microscopic entropy. These include in particular logarithmic corrections. In \cite{sen,sentwo,senthree,senfour,1204.4061} 
it has been 
shown that these logarithmic corrections can be obtained by a one-loop calculation in Euclidean quantum gravity.\footnote{A general argument showing why the logarithmic
corrections are not affected by higher loop corrections can be found in section 2.5
of \cite{1205.0971}. Even though the argument was given in the context of black
hole entropy, it holds for the partition function of quantum gravity in any background
characterized by a large overall length scale.}
The field theory background for this calculation is taken to be 
the near-horizon geometry of the black hole. 

Our problem is very similar, structurally, to the problem of computing extremal
black hole entropy. 
The CFT partition function can be regarded as the ``microscopic" result for the partition function. 
The leading large $N$ result (\ref{leadinglargen}) is the analogue of the Bekenstein--Hawking or Wald entropy. We then expect the subleading logarithmic correction 
(\ref{slog}) to be reproduced by a one-loop correction in Euclidean quantum gravity on AdS$_4 \times X_7$, as in \cite{sen,sentwo,senthree,senfour,1204.4061}. 

In this note we describe a one-loop calculation in eleven-dimensional supergravity 
(11d SUGRA) leading to a log correction of the form (\ref{slog}). The only contribution to the log corrections 
in 11d SUGRA
comes from the analysis of zero modes. This is due to the fact that in 
odd dimensions the heat kernel expansion does not contain constant terms;
an analogous situation occurs in  
the analysis of five-dimensional black holes \cite{senfour}. The only source of zero modes in the AdS$_4\times X_7$ 
background is the two-form anticommuting ghost which appears in the quantization of the SUGRA three-form. Since this does not depend on $X_7$,
this would explain the universality of the result (\ref{slog}). Our goal will be to check that the contribution from the zero modes
gives us the same coefficient of  $\log L$ that appears in \refb{slog}.

\sectiono{General strategy}

The general strategy for computing 
the logarithmic term in the 
one-loop corrections in Euclidean quantum gravity has been 
explained in \cite{sen,sentwo,senthree,senfour,1204.4061}, building on 
previous results (see for example \cite{cd,9709064}). 

The contribution of a free field to the free energy $F$ is 
divided into  
two parts: the contribution of non-zero modes, and the contribution of 
zero modes. Let us 
start with the contribution of non-zero modes. This  is  
given by the logarithm of the one-loop determinant of the kinetic operator $A$,
sans the zero modes, and takes the form
\be \label{efree}
 \mp {1\over 2} \ln {\det}' A=
\mp \frac{1}{2}\sum_{n}~'\ln {\kappa_n}
\ee
where the $'$ denotes sum over non-zero modes, 
$\kappa_n$ are the eigenvalues of the kinetic
operator, and the sign $\mp$ corresponds to 
Grassmann even/odd fields, respectively. Information about 
the spectrum of an operator $A$ is encoded in its heat kernel 
operator, defined 
as 
\be \label{edefktau}
K(\tau)=\re^{-\tau A}=\sum_{n} \re^{-\kappa_n \tau} |\phi_n \rangle \langle \phi_n |, 
\ee
where $|\phi_n \rangle$ are the corresponding eigenstates (here, for simplicity, we are assuming that the spectrum is discrete and non-degenerate; the formulae can be easily modified for more general cases). As emphasized in for example section 
2 of \cite{dt}, the heat kernel contains information about both zero and non-zero modes. 
Let us denote by $n_A^0$ the number of zero modes of the operator $A$. Then, one has the following equation 
\be
-{1\over 2} \ln {\det}' A={1\over 2} \int_\epsilon^\infty {\rd \tau \over \tau} \left( \tr \, K(\tau)-n_A^0\right)\, ,
\ee
where   
$\epsilon$ is an UV cutoff. On the other hand, 
the trace of the heat kernel has the following well-known expansion at small $\tau$, called the Seeley--De Witt 
expansion,\footnote{The coefficients $a_n$ given here were called $a_{2n}$ in
\cite{1204.4061}.}
\be \label{etracektau}
\tr K(\tau)={1 \over (4 \pi)^{d/2}}\sum_{n=0}^{\infty}  \tau^{n-d/2} \int \rd^d x  \sqrt{g}  \, a_n(x,x).
\ee
As explained in \cite{sen,sentwo,senthree,senfour,1204.4061}, one can extract from the Seeley--De Witt expansion the contribution to $\ln{\det}'A$ proportional to $\log\, L$. 
To see this, notice that since
the non-zero eigenvalues of a standard Laplace type operator
$A$ scale as $L^{-2}$,
the heat kernel is a function of 
\be
\overline \tau={\tau \over L^{2}}, 
\ee
and we can write 
\be
\label{deltas}
-{1\over 2} \ln {\det}' A= {1\over 2 }\int_{\epsilon/L^{2}} ^\infty {\rd \overline \tau \over \overline \tau} \left(\sum_{n=0}^{\infty}  {1\over  (4 \pi)^{d/2}} \overline \tau^{n-d/2} L^{2n-d} \int \rd^d x  \sqrt{g}  \, a_n(x,x) -n_A^0\right). 
\ee
The logarithmic contribution to $\ln {\det}' A$
comes from the term $n=d/2$ in 
\refb{deltas}, and we get 
\be\label{total}
-{1\over 2} \ln {\det}' A=  \left( {1\over  (4 \pi)^{d/2}}  \int \rd^d x  \sqrt{g}  \, a_{d/2}(x,x) -n_A^0 \right)   
\, \log\, L +\cdots \, ,
\ee
where $\cdots$ denote non-logarithmic contributions.
In odd--dimensional spacetimes, as it will be in our case, the coefficient $a_{d/2}$ 
vanishes,\footnote{Since AdS$_4$ is a manifold with boundary, there could be 
half integer
powers of $\tau$ in the expansion of Tr$K(\tau)$ from the boundary
(see {\it e.g.} \cite{0306138}).
However these are given by integrals of local terms over the boundary of AdS$_4$
and can be cancelled by boundary counterterms.}
and the only contribution comes from $n_A^0$. 
Combining \refb{efree} and \refb{total} we get the net
contribution to the free energy from the non-zero modes.

Note that the $\log L$ term comes from the region of integration
$\epsilon/L^2 \ll \bar \tau \ll 1$ which, in the original variable, translates to
$\epsilon \ll \tau \ll L^2$. This is the infrared region and hence is not affected
by the details of the ultraviolet cut-off $\epsilon$, which only affects the
contribution to the integral from the region $\tau\sim\epsilon$. This is important since
eleven dimensional supergravity is known to have ultra-violet 
divergences \cite{fradkin}. The reader may nevertheless worry about the
fact that the UV divergent terms could give contributions which dominate
over the logarithmic corrections, {\it e.g.} a term involving $a_n$ will give
a one loop contribution of order $L^{d-2n}$. Thus for example, the $a_0$ term,
if non-zero, would have produced a contribution of order $L^{11}$ in $d=11$.
To avoid this worry we could consider, instead of the free energy
$F$, the quantity
$(L \rd/\rd L -1) (L \rd/ \rd L -3) \cdots (L \rd/ \rd L - 11) F$. In this all polynomials
in $L$ up to order $L^{11}$ cancel, and the dominant term is proportional
to $\log L$, whose coefficient we are calculating. A similar trick was used in
\cite{1202.2070} for extracting the universal part of entanglement entropy in quantum
fleld theories.

Let us now look at the contribution coming from zero modes. As 
in \cite{sen,sentwo,senthree,senfour,1204.4061}, these arise 
due to asymptotic symmetries. To understand the dependence on $L$ of 
this integration, one evaluates the Jacobian from the coefficients of the zero modes, 
to the parameters labeling the supergroup of asymptotic symmetries. Let us 
suppose that there is a factor of 
$L^{\pm \beta_A}$ for each zero mode. Then, the total contribution to the partition
function from the zero modes is 
\be
L^{\pm \beta_A n_A^0},
\ee
and hence to the free energy is
\be
\label{zeromodecon}
\pm \, \beta_A \,  n_A^0 \log \, L \, .
\ee
Notice that our conventions for the Grassmannian case are slightly different from the ones 
used in \cite{senthree,senfour}. 

At this point, there is an important remark to be made about the computation of the number of zero 
modes $n_A^0$. Often 
in non-compact spaces the number  of zero modes, $n_A^0$, is infinite. 
Let us first suppose that our space is compact. If we call $\phi_{\ell}^{(0)}(x)$ the normalized eigenfunctions
corresponding to the zero modes, where $\ell=1, \cdots, n_A^0$, we have the equation
\be \label{ez.1}
n_A^0=  \sum_{\ell=1}^{n_A^0} \int \rd^d x  \sqrt{g} |\phi_{\ell}^{(0)}(x)|^2. 
\ee
In a non-compact space this expression is often divergent, leading to an infinite value
of $n_A^0$. 
Thus
in order to make sense of this equation, one has to find a suitable regularization of this
expression. 
In homogeneous spaces of constant curvature, like Euclidean AdS$_4$ or
AdS$_2\times\IS^2$, the sum
\be
\sum_{\ell}  |\phi_{\ell}^{(0)}(x)|^2
\ee
is a constant. 
Thus we can express \refb{ez.1} as
\be \label{ezerofin}
n_A^0= \left(\sum_{\ell} |\phi_{\ell}^{(0)}(x)|^2\right) \, \int \rd^d x  \sqrt{g}\, .
\ee
Even though the sum over $\ell$ runs over an infinite number of zero modes, 
$\sum_{\ell} |\phi_{\ell}^{(0)}(x)|^2$ is finite in cases of interest. 
Thus the divergence comes from the infinite volume of space-time,
and evaluation of \refb{ez.1} only involves 
finding a suitable regularization of the volume of space-time. 
This will be discussed for AdS$_4$ space in \S\ref{snumz}.

\sectiono{The calculation}

Our goal now is to perform a ``macroscopic" calculation of the $\log L$ correction to the free energy of 11d SUGRA on a background of the form AdS$_4 \times X_7$, and compare it to the microscopic 
prediction (\ref{slog}) from AdS/CFT. 

The most important simplification in our case is the fact that, since we are in {\it odd} dimensions, the contributions coming from the Seeley--De Witt expansion in (\ref{total}) vanish. Therefore, 
we only have to take into account the zero mode contribution (\ref{zeromodecon}). We are then led to the question of which fields lead to discrete zero modes in the background we are considering. 
Fields with zero modes play of course a crucial r\^ole in the calculations of \cite{sen, sentwo, senthree, senfour, 1204.4061}. There, the background is AdS$_2 \times \IS^2$
or AdS$_2\times$(squashed)$\IS^3$, 
and the zero modes arise from the ``exceptional" zero modes of one-forms, metric
and gravitinoes on AdS$_2$ described in \cite{chone}. 

\subsection{Expression for the logarithmic correction} \label{sexp}

In (Euclidean) AdS$_4$, the only bosonic fields which might possibly have discrete zero modes are actually two-forms, as explained in \cite{chtwo} (in general, $N$-forms have discrete zero modes on Euclidean AdS$_{2N}$). 
For fermionic fields, it can be shown that neither spinors (of spin $1/2$) nor gravitinos (of spin $3/2$) have zero modes. 

Now, there is a source of two-forms in the quantization of 11d SUGRA. This is because the quantization 
of the SUGRA three-form $C_{MNP}$ needs a generalized 
ghost field which is a Grassmannian two-form. In general, the quantization of a $p$-form $A_p$ requires $p$ generalized ghost fields $A_{p-j}$ 
which are $p-j$ forms, $j=1, \cdots, p$  \cite{siegel,thmg,ct}. 
They are Grassmann even if $j$ is even, and Grassmann odd if $j$ is odd. The action for the original $p$-form and the ghost fields, after gauge fixing, is given by
\be
S={1\over 2} \sum_{j=0}^p (p-j)! \left( A_{p-j}, (\Delta_{p-j})^{j+1} A_{p-j}\right)
\ee
where $(\cdot, \cdot)$ is the standard inner product of forms induced by the Riemannian metric, and $\Delta_k$ is the Hodge--Laplace operator acting on $k$-forms. The one-loop contribution to the free energy of the non-zero modes is then given by
\be \label{deltapre}
-\frac{1}{2} \sum_{j=0}^p (-1)^{j} (j+1)\ln \, \det \left( \Delta'_{p-j} \right)
\ee
where the $'$ indicates that we are removing the zero modes. In an 
odd-dimensional spacetime, $a_{d/2}$ vanishes and 
using \refb{total}, \refb{deltapre} 
we get the logarithmic contribution to the free energy from the non-zero modes 
to be
\be
-\sum_j (-1)^{j} (j+1) n_{\Delta_{p-j}}^0 \, \log L 
\ee
where $n_{\Delta_{p-j}}^0$ is the number of zero modes of the Hodge--Laplace 
operator $\Delta_{p-j}$. Taking into account the contribution of zero modes given in 
\refb{zeromodecon}, we obtain the general expression
\be 
\Delta F= \sum_j (-1)^{j}\left( {\beta_{p-j} }-j-1\right) n_{\Delta_{p-j}}^0 \, \log L 
\ee
for the logarithmic contribution to the free energy
of all the physical fields and ghost fields 
appearing in the quantization of a $p$ form. In our case, $n_{\Delta_{p-j}}^0$ is only different from zero when $p=3$ and $j=1$.
This gives
\be \label{egenexp}
\Delta F = -  \left(\beta_2 -2\right) \, n_{\Delta_2}^0 \, \log L\, .
\ee
Thus we have to compute $n_{\Delta_{2}}^0$ and $\beta_2$. 

Physically the $-\beta_2$ factor in \refb{egenexp} is the result of integration
over the zero modes of the 2-form field. Since the 2-form field is a ghost field,
one might wonder what it means to integrate over its zero modes. For this we
can offer the following interpretation. In theories with gauge invariance, the definition of the 
path integral involves dividing by the volume of the group of gauge transformations ${\rm vol}(\CG)$. 
In the usual Faddeev--Popov gauge fixing, 
this factor is cancelled by the ghost path integral. However, when there are zero modes in the Faddeev--Popov operator, there is only a 
partial cancellation, and after gauge-fixing the path integral still includes a factor of $1/{\rm vol}(\CH)$, where $\CH$ is the subgroup of gauge transformations 
generated by zero modes (see \cite{mmlectures}, section 3.1, for a review of this fact in the context of gauge theories, 
and \cite{vw}, section 3.4, for an example in gravity). 
In our case, since the gauge transformation 
parameters of the 3-form field are given by a 2-form, the path integral will contain in the denominator 
an integration over the zero modes of the two-form fields.
The contribution in \refb{egenexp} proportional to $\beta_2$ can then be interpreted
as the result of dividing the path integral by the integral over the zero modes of
the gauge transformation parameter. This also explains why this contribution
comes with a minus sign.

Notice that there is another potential source of zero modes, -- these could
arise if $X_7$ has a harmonic one-form so that we can get a harmonic
three-form on AdS$_4 \times X_7$ by taking the wedge product of a 
harmonic two-form on AdS$_4$ 
times a harmonic 
one-form on $X_7$. However, the $X_7$ are compact Einstein manifolds of positive curvature, and they have $b_1=0$ (see for example \cite{dnp}, page 57). Thus we 
conclude that there are no zero modes from this decomposition. 

\subsection{Calculation of the number of zero modes} \label{snumz}

To compute $n_{\Delta_{2}}^0$, we use \refb{ezerofin}.
Thus we need to calculate two quantities: 
$\sum_{\ell} |\phi_{\ell}^{(0)}(x)|^2$ and the regularized volume of AdS$_4$.
For the first quantity 
we can use 
the general result of \cite{chtwo}, which says that on 
AdS$_{M}$ and for $M/2$-forms, 
\be
\sum_{\ell} |\phi_{\ell}^{(0)}(x)|^2={1\over 2^M \pi^{M/2}}{M! \over (M/2)!}\, 
{1\over L^4}.  
\ee
For $M=4$ this gives
\be
\label{result}
\sum_{\ell} |\phi_{\ell}^{(0)}(x)|^2={3\over 4 \pi^2 L^4}. 
\ee
We can also arrive at this
result by explicitly
evaluating the left hand side at the origin of AdS$_4$. In this case only 
a few $\phi_\ell(x)$'s are non-vanishing, and by explicitly summing over the
contribution from these modes we again arrive at \refb{result}

The regularized volume of ${\rm AdS}_4$ can be calculated by standard
procedure (see for example \cite{dorn,1007.3837}) but since this forms an integral part of our
analysis, we shall review it here. For this we write the AdS$_4$ metric as
\be \label{em1}
\rd s^2 = L^2 (\rd\eta^2 + \sinh^2\eta \, \rd\Omega_3^2), \quad 0\le \eta<\infty\, ,
\ee
where $\rd\Omega_3$ is the line element on the unit 3-sphere. If we regularize the
volume of AdS$_4$ by putting a cut-off $\eta<\eta_0$ then the volume is given by
\be \label{em2}
V_{{\rm AdS}_4} = 2\pi^2 \, L^4\, \left( {1\over 24} \, \re^{3\eta_0} -{3\over 8} \re^{\eta_0}
+{2\over 3} + \OO(\re^{-\eta_0})\right)\, .
\ee
On the other hand the radius of curvature of the boundary 3-sphere at
$\eta=\eta_0$ is given by
$R = L\sinh\eta_0 = L (\re^{\eta_0} - \re^{-\eta_0})/2$. Thus the terms in \refb{em2}
proportional to $\re^{3\eta_0}$ and $\re^{\eta_0}$ can be expressed as 
polynomials in $R$ up to order $1/R$ corrections and hence can be cancelled by boundary
counterterms. As a result we are left with the regularized volume
\be \label{evolads}
{\rm vol}({\rm AdS}_4)= {4 \pi^2 L^4 \over 3}. 
\ee
Notice that this is the same regularized value which leads to the successful test of the leading term (\ref{leadinglargen}). 
Substituting \refb{result} and \refb{evolads} into \refb{ezerofin} we get
\be \label{enzero}
n_{\Delta_2}^0=1. 
\ee

\subsection{Calculation of $\beta_2$}

We now compute $\beta_2$. To do this, we 
proceed as in \cite{sen,sentwo,senthree,senfour,1204.4061}.
The path integral measure over the two-form $B_{\mu \nu}$ in $D$ dimensions is normalized as 
\be
\int [ {\cal D} B_{\mu \nu} ]\exp\left[-\int \rd^D x\sqrt{g}g^{\mu \nu}g^{\alpha \beta}  B_{\mu \alpha} B_{\nu\beta}\right]=1.
\ee
The metric on AdS$_4\times X_7$
can be written as $g_{\mu \nu}=L^{2}g_{\mu \nu}^{(0)}$, where  $g_{\mu \nu}^{(0)}$ is an $L$-independent metric. The normalization becomes
\be
\int [ {\cal D} B_{\mu \nu} ]\exp\left[-L^{D-4}\int \rd^D x\sqrt{g^{(0)}}  
g^{(0)\mu \nu} g^{(0)\alpha \beta}  B_{\mu \alpha} B_{\nu\beta}\right]=1.
\ee
Hence the correctly normalized integration measure corresponds to an integration
\be \label{ebrange}
\prod_{x, (\mu \nu)} \rd\left( L^{ D/2-2 }B_{\mu \nu}(x) \right).
\ee
On the other hand, the zero modes of the $B_{\mu \nu}$ field are associated with the usual gauge transformation of two-forms,
\be \label{ebrange2}
\delta  B_{\mu \nu} \propto \partial_{\nu} \theta_{\mu}-\partial_{\mu} \theta_{\nu}\, ,
\ee
but with non-normalizable $\theta_\mu$ so that these are not pure gauge deformations. 
Now, since we are using a coordinate system in which the metric takes the form
$L^2 g^{(0)}_{\mu\nu}$, the range of coordinates is independent of $L$. 
Since in any coordinate system we expect $\int \theta_\mu \rd x^\mu$ to have
$L$ independent periods, in the coordinate system used here, in which
$x^\mu$'s have $L$ independent range, 
the $\theta_\mu$'s should have $L$ independent integration range.
Eq. \refb{ebrange2} now shows that the integration over each
$B_{\mu\nu}$ zero mode has an $L$-independent integration range, but due to the 
$L^{D/2-2}$ factor in the measure in \refb{ebrange} 
it  gives a factor of $L^{D/2 -2}$.
Thus we have\footnote{Since $B_{\mu\nu}$ are Grassmann variables  
the factor is actually $L^{-(D/2 -2)}$ but the extra minus sign has already been taken
into account in \refb{egenexp}.}
\be \label{ebeta2}
\beta_2={D\over 2}-2 = {7\over 2}
\ee
for $D=11$.

Since the above result depends crucially on the result for the range of integration
of the zero mode of $B_{\mu\nu}$ we shall now elaborate on this further
in the context of a compact manifold,  regarding the $B_{\mu\nu}$ as the
gauge transformation parameters of the 3-form field. For this
let us consider, instead of AdS$_4\times X_7$, a compact space with metric
$L^2 g^{(0)}_{\mu\nu}$ where $g^{(0)}_{\mu\nu}$ is $L$ independent. In this case
the zero modes of $B_{\mu\nu}$ are harmonic 2-forms, which can be represented
locally as $\rd\Lambda$ for some one-form $\Lambda$, but this one-form is not globally
defined. This is analogous to the 2-form zero modes on AdS$_4\times X_7$ which
are locally of the form $\rd\Lambda$ but the one-form $\Lambda$ is not
normalizable. Now returning to the compact case we see that if we regard the
two-forms as the gauge transformation parameters of the three-form fields,
then the integral of the two form over a 2-cycle of the manifold is a global symmetry
transformation parameter, and the corresponding conserved charged is the
winding number of the M2-brane on this 2-cycle. Since the latter is quantized in
integer units, the integral of the 2-form over the 2-cycles will have period $2\pi$. 
Since $g^{(0)}_{\mu\nu}$ and
hence the coordinate system we have used has
no explicit dependence on $L$, this shows that the zero modes
of the $B_{\mu\nu}$ fields have
$L$ independent integration range. 

It is also worth noting that if instead of the 2-form field we had zero modes of the
metric -- as in the case of \cite{sen,sentwo,senthree,senfour,1204.4061} -- then
the result would be different. In this case the metric zero modes would be associated with
diffeomorphisms with non-normalizable transformation parameters $\xi^\mu(x)$, 
and the analog of \refb{ebrange2} would be $\delta g_{\mu\nu} =\nabla_\mu \xi_\nu
+\nabla_\nu \xi_\mu$. However now the natural variables which have $L$ independent
range are the transformation parameters $\xi^\mu$, and so when we lower the
index with the metric $g_{\mu\nu}$, we get a factor of $L^2$ in the range of integration
over the metric zero modes \cite{sen,sentwo,senthree,senfour,1204.4061}.
Again the validity of this argument can be checked using the example of a compact
manfold.
We take the familiar 
example of a square torus $\IT^2$ with metric $\rd s^2 = L^2 (\rd x^2 + \rd y^2)$
and take $x, y$ to have period 1. Now consider a diffeomorphism $y\to y + a x$, 
$x\to x$, 
under which the metric is deformed to $L^2 (\rd x^2 + (\rd y + a \rd x)^2)$. Since this
diffeomorphism does not preserve the periodicity in $x$ and $y$, it is not an allowed
diffeomorphism and hence generates a genuine deformation of the metric. Thus this
is analogous to non-normalizable diffeomorphisms in the non-compact case. But
for $a=1$ the periodicity in $x$ and $y$ is preserved showing that 
$a=1$ is the same as $a=0$. 
Hence the parameter $a$
has period 1, independent of $L$, as we expect on general grounds. 
Note however that since
under this deformation the metric changes by order $L^2 \delta a$, 
the range of integration
over the metric zero mode is of order $L^2$, as predicted from our general
arguments.

\subsection{Logarithmic correction to the free energy}

Using \refb{egenexp},
\refb{enzero} and \refb{ebeta2} we see that the logarithmic correction 
to the free energy is given by
\be
-\left(\beta_2-2\right) \log L = -{3 \over 2} \log L
\ee
which precisely matches (\ref{slog}). 

\sectiono{Conclusions and outlook}

In this note we have shown that the logarithmic correction to the three-sphere partition function, in a large class of three-dimensional 
CFTs generalizing ABJM theory, can be computed by doing a one-loop calculation in the dual eleven-dimensional supergravity on AdS$_4 \times X_7$. 
This can be regarded as a generalization of the 
program for calculating logarithmic corrections to the black hole entropy developed in \cite{sen,sentwo,senthree,senfour,1204.4061}, and provides a non-trivial test of the 
AdS/CFT correspondence at next-to-leading order in the $1/N$ expansion.
We have found that this correction is due only to zero modes, more precisely, to the zero mode of a ghost two-form appearing in the quantization of the three-form field of supergravity. 
This explains its universality: on the field theory side, the correction is independent 
of  
the data of the CFT, and on the supergravity side, it is independent of the seven-dimensional 
manifold $X_7$. 

The computation we have done here can be extended in various directions. For example, it would be interesting to reproduce the logarithmic shift in the type IIA string picture obtained 
by dimensional reduction of the eleven-dimensional supergravity backgrounds considered in this paper, which leads to 
backgrounds of the form AdS$_4\times X_6$. In this case there will be contributions from 
both non-zero and zero-modes, and the answer depends in principle on the details of the six-dimensional manifold $X_6$ appearing in the compactification. Another interesting extension 
concerns the study of logarithmic corrections for type IIA string theory on the backgrounds of the form AdS$_6\times X_4$ found in \cite{brg}. In this case the partition function can be also computed in the CFT side, and it agrees with the gravity dual at large $N$ \cite{jp}, so one might try to compare the logarithmic corrections.

\section*{Acknowledgements}
We would like to thank Matthias Gaberdiel, Juan Maldacena, Ruben Minasian, Gregory Moore, Binata Panda,
Martin Rocek and Arkady Tseytlin 
for useful conversations. M.M. would like to thank the 2011 Simons Workshop for hospitality in the 
preliminary stages of this work. The work of A.G. and M.M. is supported by the Fonds National Suisse under subsidies 200020-126817 and 
200020-137523. The work of S.B. and A.S. is supported in part
by  the 
project 11-R\&D-HRI-5.02-0304. The work of A.S. is also supported
in part by the J. C. Bose fellowship of 
the Department of Science and Technology, India.

\end{document}